\documentclass[fleqn,12pt,twoside]{article}
\usepackage[headings]{espcrc1}

\readRCS
$Id: espcrc1.tex,v 1.2 2004/02/24 11:22:11 spepping Exp $
\usepackage{graphicx}
\usepackage[figuresright]{rotating}

\newcommand{\AmS}{{\protect\the\textfont2
  A\kern-.1667em\lower.5ex\hbox{M}\kern-.125emS}}

\title{Measurement of event-by-event fluctuations 
       and order parameters in PHENIX}

\author{T. Nakamura\address[MCSD]{Hiroshima University, 
        1-3-1 Kagamiyama, Higashi-Hiroshima 739-8526, Japan} 
	for the PHENIX Collaboration
        \thanks{For the full list of PHENIX authors and acknowledgments, 
	        see Appendix 'Collaborations' of this volume.}}
       
\runtitle{Measurement of event-by-event fluctuations and order parameters in PHENIX}
\runauthor{T. Nakamura for the PHENIX Collaboration}

\begin{document}

\maketitle

\begin{abstract}
We present the latest results on event-by-event fluctuations of charged particle 
multiplicity in Au+Au and Cu+Cu collisions at $\sqrt{s_{NN}} = 200$ GeV and $62.4$ GeV 
measured by PHENIX experiment at RHIC. 
The two particle correlation length to discuss order parameters, which can be extracted from 
scale dependence of the fluctuations, is supposed to be sensitive to the critical points 
of QCD phase transition. 
The obtained correlation length with respect to the pseudo rapidity gap indicates the 
power law behavior as a function of the number of participant nucleons for Au+Au collisions 
at $\sqrt{s_{NN}} = 200$ GeV. 
\end{abstract}

\section{Introduction}
Event-by-event fluctuations on charged particle multiplicity and transverse momentum 
have been studied to understand the properties of QCD phase in heavy-ion collisions. 
Several candidates of order parameters in QCD phase, for example correlation length 
\cite{Carruthers} and specific heat \cite{Korus,Mekjian}, can be extracted from these observables. 
It enables us to characterize the critical point of the phase transition by surveying 
the behavior of order parameters as a function of temperature or energy density of the system. 
This is one of the basic methods in the condensed matter physics. 
We have measured charged particle multiplicity fluctuations in Au+Au and Cu+Cu collisions 
at $\sqrt{s_{NN}} = 200$ GeV and $62.4$ GeV by using the PHENIX central arm spectrometer 
located at mid rapidity region ($|\eta| < 0.35$) \cite{PHENIX}. 
The extraction of two particle correlation length has been performed from the multiplicity 
fluctuations measured in various sizes of pseudo rapidity gaps by using an integral of 
two particle correlation function.

\section{Charged particle multiplicity distribution}
In an approximate sense, multiplicity distributions of charged particles produced 
in several high energy collisions such as $e^{+}e^{-}$, p+p collisions \cite{Dremin}
and nucleus-nucleus collisions \cite{E802} are well agree with the negative binomial 
distributions (NBD) defined as 
\begin{equation} 
\label{eqn:NBDE}
P_n^{(k)} = \frac{\Gamma(n+k)}{\Gamma(n-1)\Gamma(k)} \left(\frac{\mu/k}{1+\mu/k}\right) \frac{1}{(1+\mu/k)},
\end{equation}
where $\mu$ indicates the average multiplicity.
The finite $k$ value in Eq. (\ref{eqn:NBDE}) describes the deviation from the 
Poisson distribution, because NBD corresponds to the Poisson distribution in 
the limit of infinity for the parameter $k$. 
In this respect, NBD $k$ parameter represents the multiplicity fluctuation. 

\begin{figure}[h]
  \begin{minipage}{80mm}
    \begin{center}
      \includegraphics[scale=0.42]{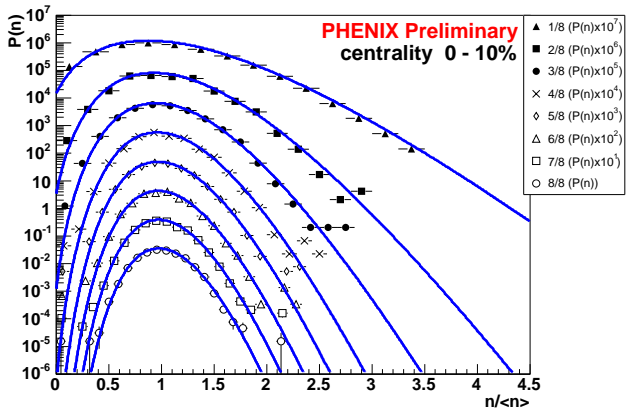}
    \end{center}
  \end{minipage}
  \hspace{5mm}
  \begin{minipage}{75mm}
    \begin{center}
      \caption{
	Multiplicity distributions for the various pseudo rapidity gaps
	measured by $0-10\%$ centrality in Au+Au at $\sqrt{s_{NN}} = 200$ GeV.
	The horizontal axis is normalized by the average multiplicity.
	Pseudo rapidity gaps are divided into several bins as indicated.
	Each distribution is scaled by the factor indicated.
	Solid lines are fit results with NBD.
      }
      \label{fig:mul}
    \end{center}
  \end{minipage}
\end{figure}

We have confirmed, that the multiplicity distributions measured by PHENIX agree with the NBD 
for all collision centralities. 
Fig. \ref{fig:mul} shows the multiplicity distributions measured in the subdivided acceptances of
the pseudo rapidity and the NBD fits to the data in $0 - 10\%$ centrality.
The collision centralities are defined by the forward detectors in PHENIX \cite{PHENIX}. 
These are independent subsystems from the central arm detectors. 
The geometrical acceptance is divided into 1/8 through 8/8 of the total acceptance 
of $\Delta\eta = 0.7$. 
For the azimuthal angle, the acceptance is selected as $\Delta\phi < 1/2\pi$.
Events collected without magnetic field are used to enhance charged particles with low 
momenta.

\vspace{-10mm}
\vspace{-10mm}

\begin{figure}[h]
  \begin{minipage}{80mm}
    \begin{center}
      \includegraphics[scale=0.36]{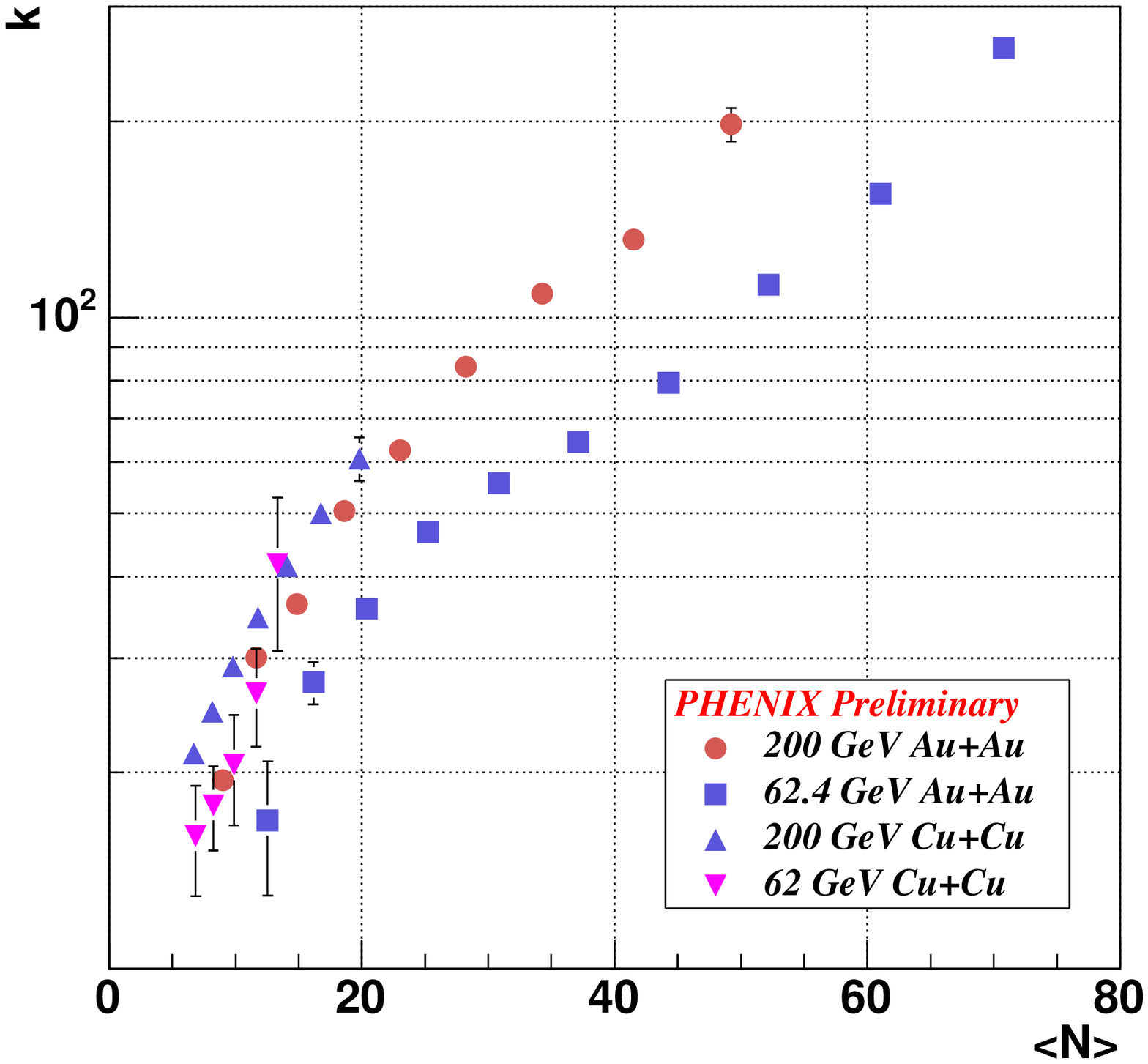}
    \end{center}
  \end{minipage}
  \begin{minipage}{80mm}
    \begin{center}
      \includegraphics[scale=0.36]{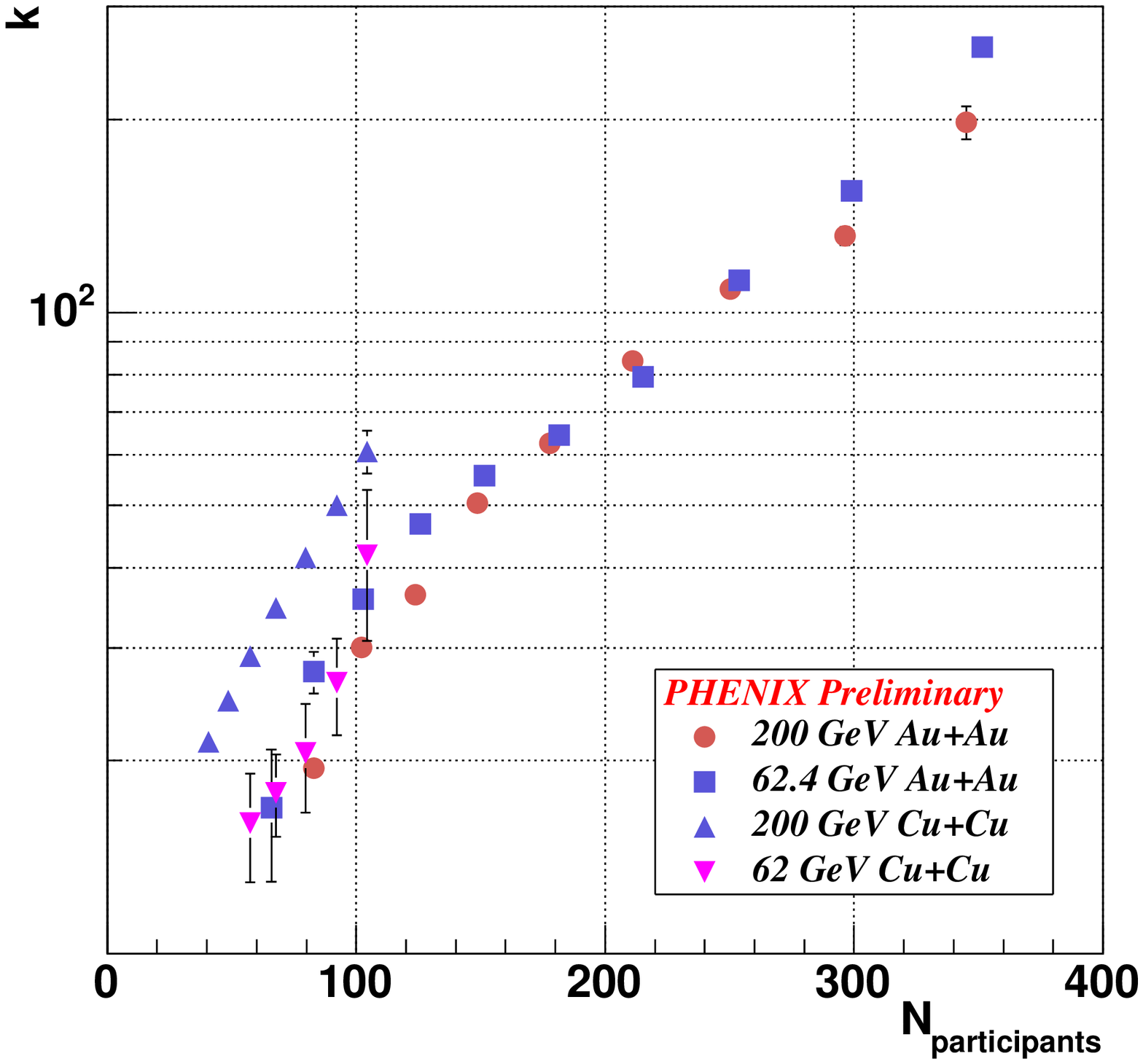}
    \end{center}
  \end{minipage}
  \vspace{-11mm}
  \caption{
    Measured NBD $k$ parameters as a function of the average multiplicity (left) 
    and number of participant nucleons (right) at the transverse momentum
    range from 0.2 to 2.0 GeV/c. Markers correspond to the collision systems 
    and energies as indicated.
  }
  \label{fig:jeff}
\end{figure}

Fig. \ref{fig:jeff} shows the measured NBD $k$ parameters as a function of average 
multiplicity and the number of participant nucleons in Au+Au and Cu+Cu collisions at
$\sqrt{s_{NN}} = 200$ GeV and $62.4$ GeV.
These are measured with the magnetic field.
One can find that the NBD $k$ parameters are not scaled by the average multiplicity but 
scaled by the number of participant nucleons for the different energies in Au+Au collisions.

\vspace{-10mm}

\vspace{-8mm}
\begin{figure}[h]
  \begin{minipage}{80mm}
    \begin{center}
    \includegraphics[scale=0.36]{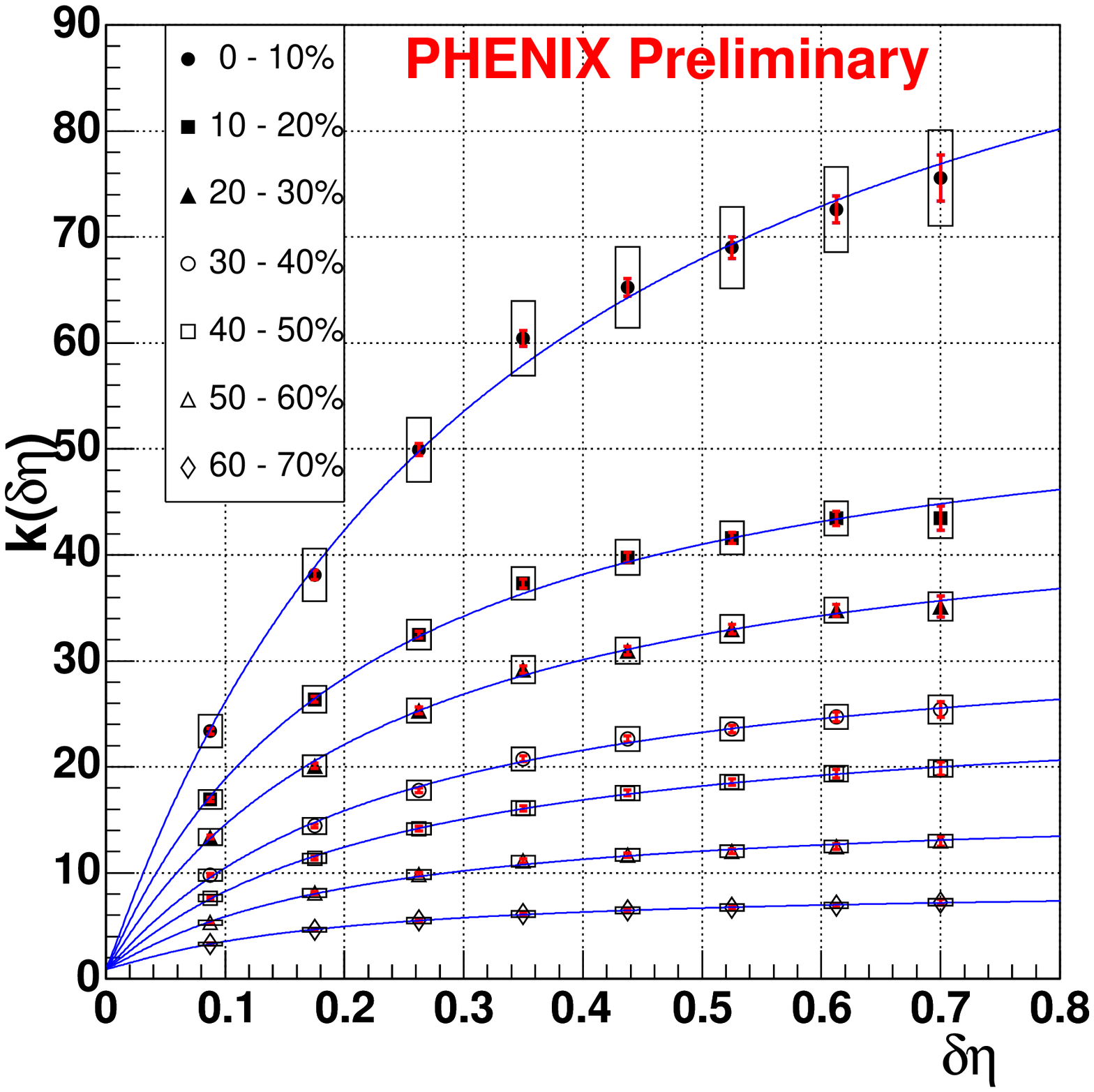}
    \end{center}
  \end{minipage}
  \begin{minipage}{80mm}
    \begin{center}
    \includegraphics[scale=0.36]{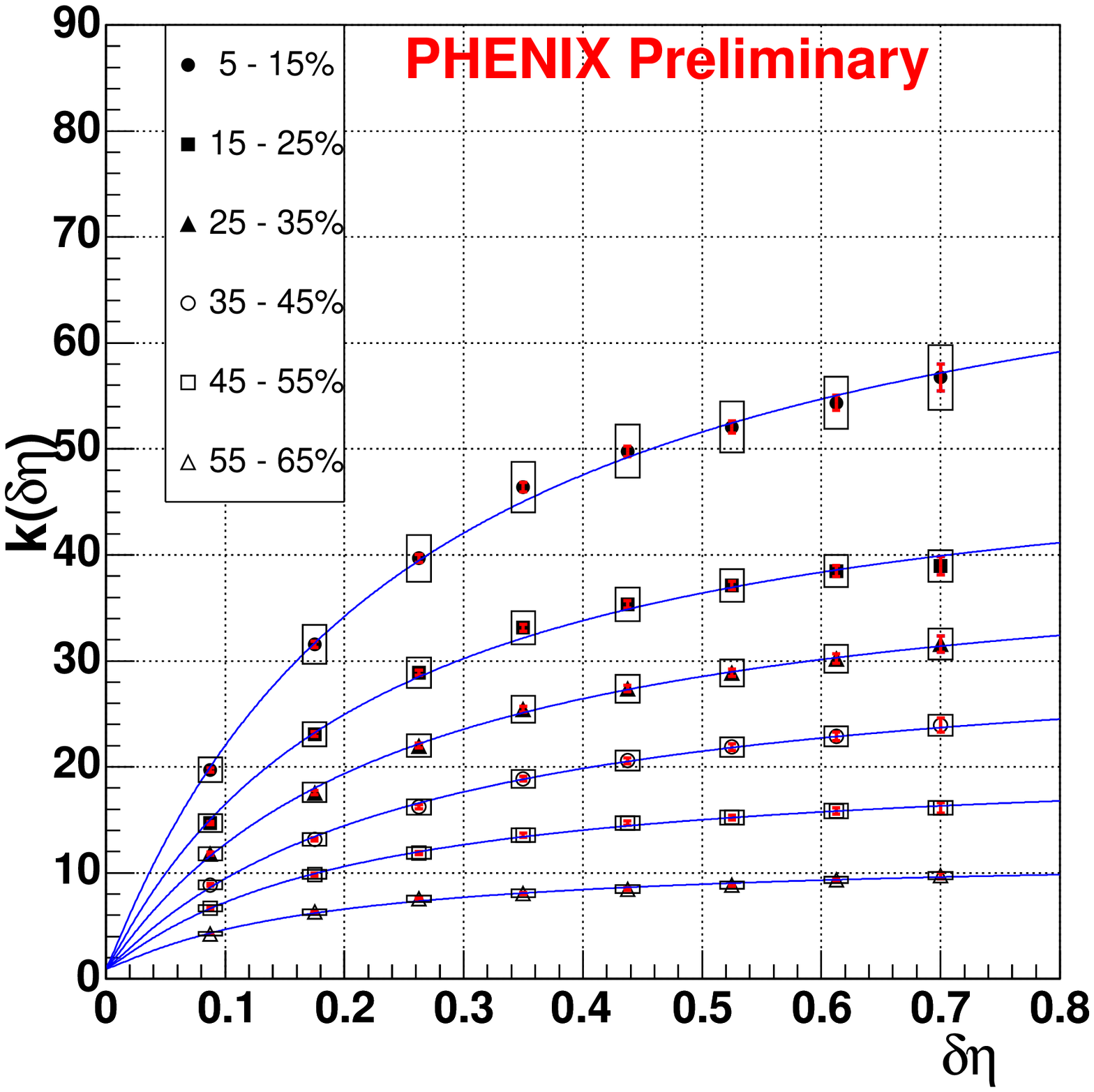}
    \end{center}
  \end{minipage}
  \vspace{-12mm}
  \caption{
    Corrected NBD $k$ parameters as a function of pseudo rapidity gap
    measured in $0\%$ through $70\%$ (left) and $5\%$ through $65\%$ (right) 
    collision centrality with $10\%$ interval respectively. Error bars indicate
    statistical errors and boxes indicate systematic errors.
    Solid lines are fits to Eq. (\ref{eqn:length_phenix}).
  }
  \label{fig:kdeta}
\end{figure}
\vspace{-12mm}

\section{Extraction of correlation length}
\begin{figure}[h]
  \begin{minipage}{80mm}
    \begin{center}
      \includegraphics[scale=0.38]{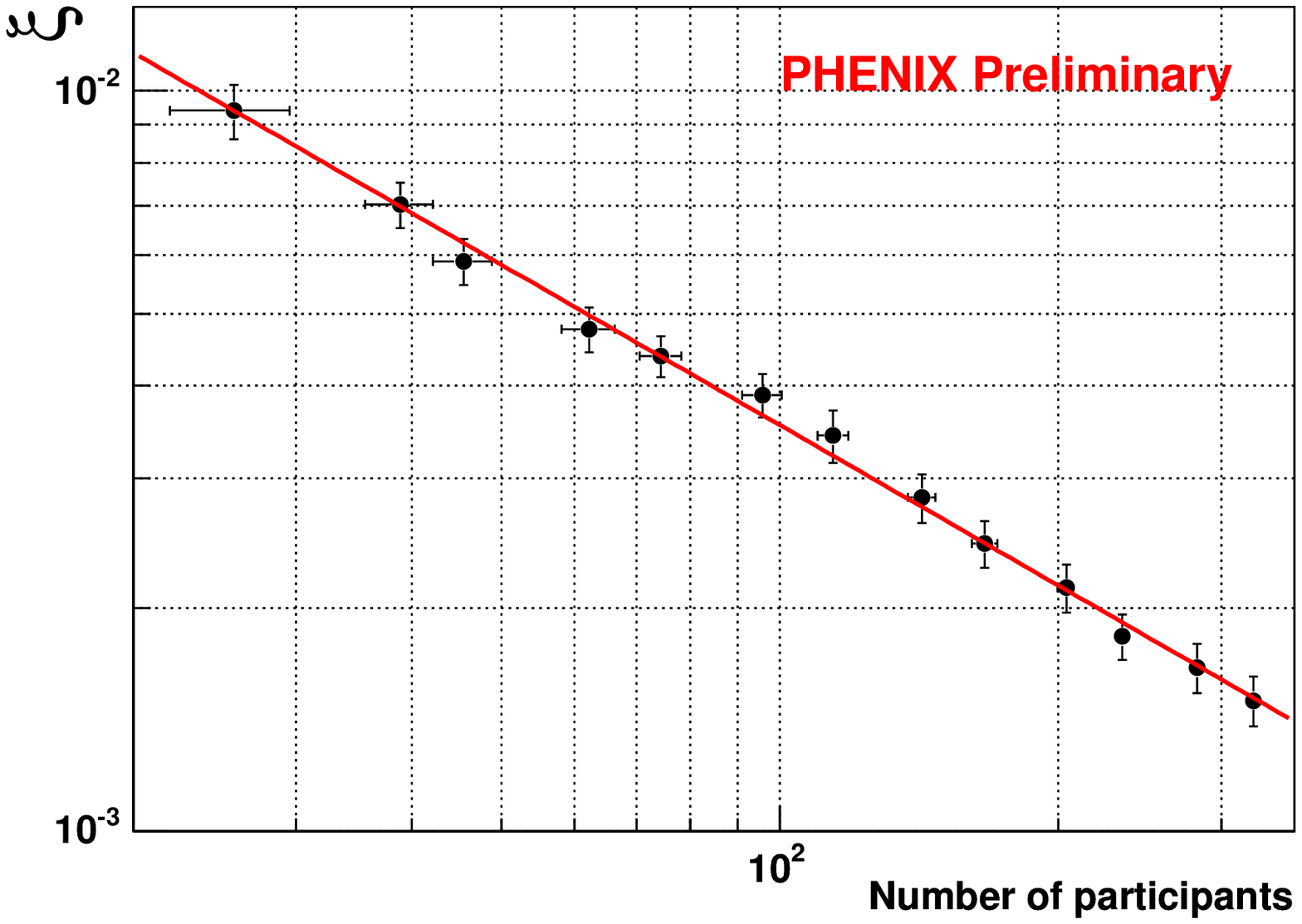}
    \end{center}
  \end{minipage}
  \begin{minipage}{80mm}
    \begin{center}
      \includegraphics[scale=0.38]{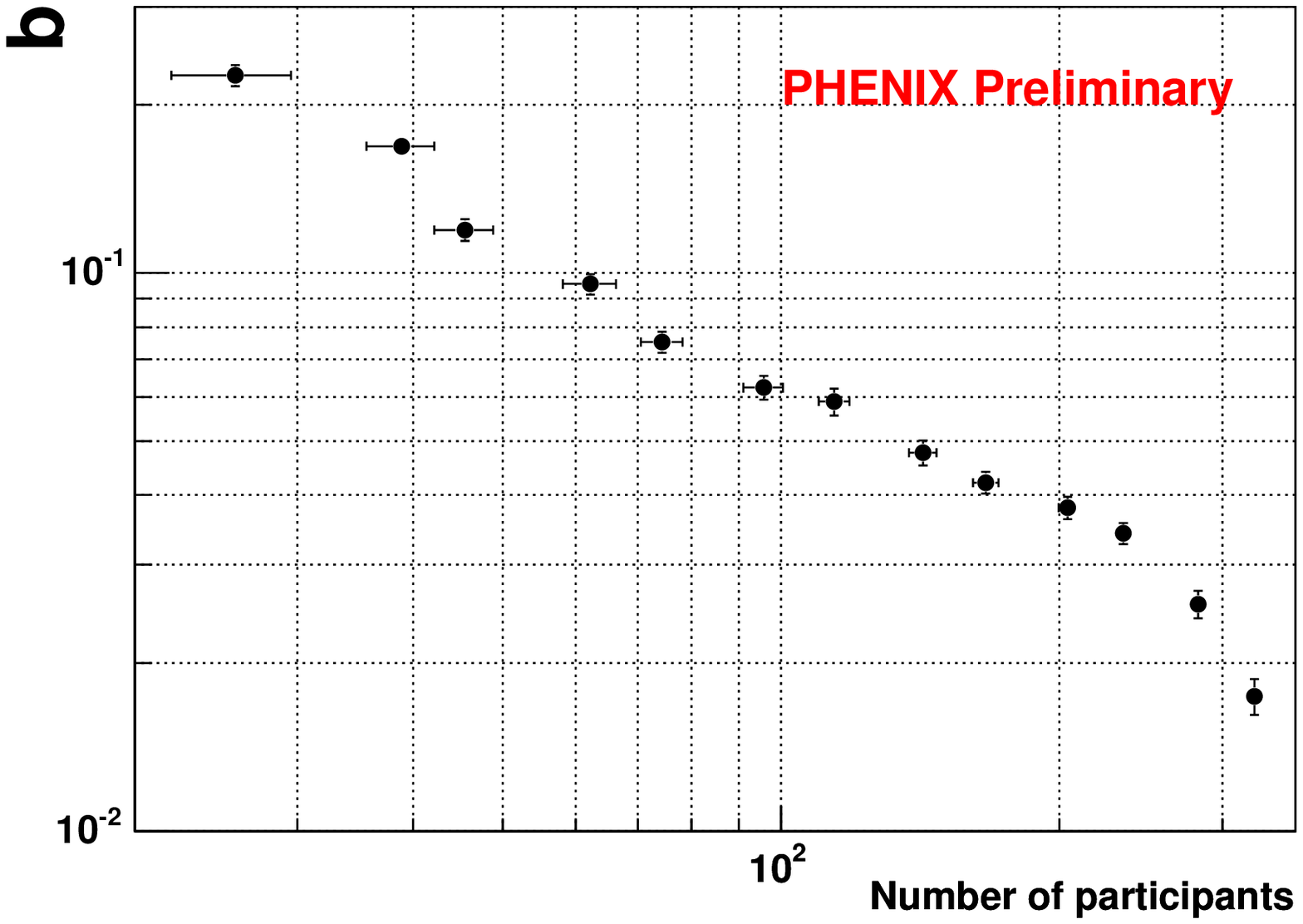}
    \end{center}
  \end{minipage}
  \vspace{-12mm}
  \caption{
    Extracted parameters $\xi$ (left) and $b$ (right) in Eq. (\ref{eqn:length_phenix})
    as a function of the number of participant nucleons. Solid line in the left figure
    is a fit with one exponent. Error bars indicates the fitting errors.
  }
  \label{fig:par}
\end{figure}

The correlation length $\xi$ in the unit of pseudo rapidity can be extracted 
from the $\delta\eta$ dependence of NBD $k$ parameters by assuming a
correlation function,
\begin{equation}
\label{eqn:c2}
C_{2}(y_{1},y_{2})=\rho_{2}(y_{1},y_{2})-\rho_{1}(y_{1})\rho_{1}(y_{2}),
\end{equation}
where $\rho_{2}$ and $\rho_{1}$ are the inclusive two particle and the single particle density respectively.
We introduced a normalized correlation function defined and parametrized as 
\begin{equation}
\label{eqn:r2}
R_{2}(y_{1},y_{2})\equiv\frac{C_{2}(y_{1},y_{2})}{\rho_{1}(y_{1})\rho_{1}(y_{2})}=ae^{-|y_{1}-y_{2}|/\xi}+b,
\end{equation}
where $\xi$ is the two particle correlation length and $a$ is the strength of two particle correlations.
Eq. (\ref{eqn:r2}) dose not have a power term with respect to the pseudo rapidity gap size, 
which is generally introduced to discuss the correlation length around the phase transitions, 
but it is a reasonable formulation for the one dimensional observable \cite{Fisher}.
The two particle correlation function is usually defined by assuming that particles are in 
completely separated phase-space ($y_{1} \neq y_{2}$) implicitly. 
However, the function can not be applicable when particles are found in the same 
phase-space gap ($y_{1} = y_{2}$). 
We introduced the constant parameter $b$ to take this into account \cite{Stanley}. 
The correlation length can be obtained by the integration of correlation function $C_{2}$ 
by assuming the constant single particle density, which is defined as
\begin{equation}
\label{eqn:k2}
F_{2}-1 = K_{2} = \frac{\int^{\delta\eta}C_{2}dy_{1}dy_{2}}{\int^{\delta\eta}\rho_{1}(y_{1})\rho_{1}(y_{2})dy_{1}dy_{2}},
\end{equation}
where $F_{2}$ is the second order normalized factorial moment and $K_{2}$ is 
named as the factorial cumulant \cite{Dremin}. 
Mathematically, $K_{2}$ coincides with the inverse of NBD $k$ parameter.
In the case of the correlation function parametrized in Eq. (\ref{eqn:r2}), 
a relation between the NBD $k$ parameter and the integrated two particle correlation function is obtained,
\begin{equation}
\label{eqn:length_phenix}
\frac{1}{k(\delta\eta)} = F_{2}-1 = K_{2} = \frac{2a\xi^{2}[\delta\eta/\xi-1+e^{-\delta\eta/\xi}]}{\delta\eta^{2}} + \frac{b}{2}.
\end{equation}
Fig. \ref{fig:kdeta} shows measured NBD $k$ values, which are corrected by the MC simulation,
as a function of pseudo rapidity gap sizes
in the Au+Au collisions at $\sqrt{s_{NN}} = 200$ GeV.
The correlation lengths are extracted by the fits to Eq. (\ref{eqn:length_phenix})
by fixing the parameter $a$ at 1. 
Fig. \ref{fig:par} indicates the extracted parameters $\xi$ and $b$ as a function 
of the number of participant nucleons. Correlation length indicates a power law behavior in the relation 
with the number of participant nucleons.

\vspace{-8mm}
\vspace{-12mm}

\section{Summary}
We measured event-by-event fluctuations for the charged particle multiplicity in Au+Au and Cu+Cu
collisions at $\sqrt{s_{NN}} = 200$ GeV and $62.4$ GeV. The measured charged particle distributions
by PHENIX agree with the NBD as well as in the past experiments. NBD $k$ parameters are not 
scaled by the average multiplicity but scaled by the number of participant nucleons or system size
in the Au+Au collisions for the different collision energies. The scale dependence of NBD $k$ parameters
for the various pseudo rapidity gap sized are presented. The extracted two particle correlation lengths 
from the scale dependence have a power law behavior as a function of the number of participant nucleons.

\end{document}